# Hot Explosions in the Cool Atmosphere of the Sun

H. Peter,[1]* H. Tian,[2] W. Curdt,[1] D. Schmit,[1] D. Innes,[1] B. De Pontieu,[3,7] J. Lemen,[3]
A. Title,[3] P. Boerner,[3] N. Hurlburt,[3] T. D. Tarbell,[3] J. P. Wuelser,[3]
J. Martínez-Sykora,[3,4] L. Kleint,[3,4,5] L. Golub,[2] S. McKillop,[2] K. K. Reeves,[2] S. Saar,[2]
P. Testa,[2] C. Kankelborg,[6] S. Jaeggli,[6] M. Carlsson,[7] V. Hansteen[7]

[1]Max Planck Institute for Solar System Research, 37077 Göttingen, Germany.
[2]Harvard-Smithsonian Center for Astrophysics, 60 Garden Street, Cambridge, MA 02138, USA.
[3]Lockheed Martin Solar and Astrophysics Laboratory, 3251 Hanover St., Org. ADBS, Bldg. 252, Palo Alto, CA 94304, USA.
[4]Bay Area Environmental Research Institute, 596 1st St West, Sonoma, CA 95476, USA.
[5]NASA Ames Research Center, Moffett Field, CA 94305, USA.
[6]Department of Physics, Montana State University, Bozeman, P.O. Box 173840, Bozeman, MT 59717, USA.
[7]Institute of Theoretical Astrophysics, University of Oslo, P.O. Box 1029, Blindern, NO-0315 Oslo, Norway.
*Correspondence to:  peter@mps.mpg.de

**The solar atmosphere was traditionally represented with a simple one-dimensional model. Over the past few decades, this paradigm shifted for the chromosphere and corona that constitute the outer atmosphere, which is now considered a dynamic structured envelope. Recent observations by IRIS (Interface Region Imaging Spectrograph) reveal that it is difficult to determine what is up and down even in the cool 6000-K photosphere just above the solar surface: this region hosts pockets of hot plasma transiently heated to almost 100,000 K. The energy to heat and accelerate the plasma requires a considerable fraction of the energy from flares, the largest solar disruptions. These IRIS observations not only confirm that the photosphere is more complex than conventionally thought, but also provide insight into the energy conversion in the process of magnetic reconnection.**

The energy produced in the core of the Sun by the fusion of hydrogen into helium is transported toward the surface first by radiation, and then by convection. The layer where the photons become free to escape defines the visible surface of the Sun. The atmosphere of the Sun above the surface was traditionally described as one-dimensionally stratified. Moving outward from the photosphere, the innermost layer, the temperature drops before rising again slightly in the middle layer, the chromosphere. When the outgoing energy – transported by a heating mechanism that is not yet fully understood – can no longer be buffered by radiative loss and hydrogen ionization, the temperature rises steeply. This transition marks the boundary of the corona, the outermost layer, which is brilliantly visible to the naked eye in a total solar eclipse. Semi-empirical models represent this simplified one-dimensional stratification well (*1*). However, more advanced observations and models have established that the outer atmosphere (chromosphere and corona) is highly structured and dynamic (*2,3,4*). Modern models of the solar atmosphere also take into account its three-dimensional dynamic nature (*5,6,7*). In the photosphere, these models deviate only mildly from the average temperature and density stratification in the semi-empirical models, typically by a factor of two or less (*8*). In particular, no evidence has been found so far, either in models or observations, for pockets of hot gas in the cool photosphere, i.e., an inversion of the temperature structure. Here we exploit the extensive temperature coverage of the IRIS spectra (*9*) to show that very significant deviations of the temperature structure exist even in the dense (upper) photosphere: small pockets of plasma that reach temperatures of almost 100,000 K appear to be embedded in the 4000 K (or cooler) photosphere. This finding has consequences for our understanding of the structure and energetics close to the solar surface and sheds new light on the conversion of magnetic energy into heat and flows in plasma physics in general. This underlines the value of the low solar atmosphere as a laboratory in particular for partly ionized collisional plasmas. Such conditions are common in astrophysics, such as in molecular clouds or proto-planetary discs.

The observations presented here concentrate on an emerging active region (see supplementary material, SM, S1). In such a region new sunspots that host magnetic fields up to 0.3 T appear in the solar photosphere. The magnetic flux that emerges through the surface gives rise to magnetic activity in which magnetic energy is converted into other forms, most noticeably internal energy, i.e., heating the plasma, and kinetic energy through the acceleration of plasma. This results in heating and mass loading of coronal loop structures, which follow magnetic field lines. These can be seen in the emerging active region (Fig. 1) connecting regions of opposite magnetic polarities. Small round brightenings are also visible in a spectral line image (Si IV, Fig. 1A) that samples gas at ~80,000 K equilibrium temperatures; these features are the pockets of hot gas in the cool photosphere.

The temporal evolution in the IRIS slit-jaw images shows that these roundish features are transient and have a lifetime of ~5 min. This is long enough for Si IV to reach ionization equilibrium (SM S2). To some extent, these brightenings share properties of Ellerman bombs (*10,11*): the latter manifest themselves as brightenings in the wings of the Hα line at 6563 Å and in UV continua (*11*). They are found in emerging active regions and are believed to occur in the photosphere (*12,13*) in response to reconnection (*14,15*), with no signature in coronal emission (above $10^6$ K) (*16*). Like Ellerman bombs, the events we report occur in regions on the surface where magnetic flux of opposite polarities converges and cancels (Fig. 1, SM S2). However, observations and models show that temperatures in Ellerman bombs are well below 10,000 K (*17, 18, 19*), which contrasts the hot pockets we find. As yet, there are insufficient IRIS data overlapping with ground-based Hα measurements to determine whether the events we report here are identical to or different from the well-studied Ellerman bombs. Because of this missing observational link and the difference in temperature, we refer to these simply as bombs (and in Fig. 1A, as bombs 1-4).

The principal tool for analyzing the thermal structure of the upper atmosphere is spectroscopy at ultraviolet wavelengths, where IRIS provides unprecedented spatial, temporal, and spectral resolution. We first show an average spectrum for the quiet plage region surrounding the emerging active region with the main lines of C II, Si IV, Fe XII, Cl I, O IV and Mg II (Fig 2; see SM S1, Table S1 for a complete list of lines). The line profiles generally conform to a single Gaussian, where they form under optically thin conditions. The lines of Mg II (and to a lesser degree, those of C II) show self-absorption in the center due to large opacity.

In contrast to the average spectrum, the spectra from the bombs are very peculiar – they are fundamentally different from any other individual spectrum from the active region and surrounding plage area outside the bomb regions (Fig. 3; other cases described in SM S3). The line profiles of the Si IV doublet at 1394 Å and 1403 Å are striking for their clear double-peaked profile. The ratio of these lines under optically thin conditions should be 2; here it is 1.95. Thus, the dip in the middle is not caused by self-absorption. The profiles therefore indicate a bi-directional flow with a speed of about ±75 km/s toward and away from the observer (corresponding to the Doppler shifts of the two components). Because this observation was acquired close to the center of the solar disk, this indicates a predominantly vertical bi-directional flow.

The simple fact that Si IV is visible and remains visible for longer than the ionization and recombination times shows that the plasma is heated to (at least) about 80,000 K, the line formation temperature of Si IV (SM S2). Our IRIS observations reveal that absorption lines from singly ionized species (Fe II and Ni II) are found in these Si IV profiles (Fig. 3). These absorption features are typically blueshifted by ~5 km/s (SM Table S1). The presence of absorption features superimposed on emission lines implies that cool material is stacked on top of hot material.

These observations are compatible with the following scenario (Fig. 4): Similar to another observationally motivated scenario (*17*) and three-dimensional models for Ellerman bombs (*18,19*), serpentine magnetic field lines form in the process of flux emergence, which produce magnetic dips in the photosphere (*20,21*). The corresponding opposite polarities at the bomb locations are clear (Fig. 1 and Movie 1;see also SM S2). The resulting U-loop is dragged down by the mass it accumulated (*22,23*), and in the upper part of the U shape, the magnetic field reconnects. In response to the explosion caused by reconnection, the plasma in the photosphere is heated to almost 100,000 K, and a bi-directional flow channeled by the magnetic field is initiated. A density analysis based on the IRIS observations of O IV and Si IV shows that the bombs form in the photosphere (SM S4). The overlying (pre-existing) cool chromosphere is slowly shifted upwards by the flux emergence (*24,25*), which explains the slight blueshift of the absorption lines from the overlying cool layer. In this scenario, the presence of hot dense pockets of gas deep in the atmosphere implies that the normal temperature stratification (hot transition region gas above the chromosphere) is turned upside down. Because the density in these bombs is so high it might be that the ionization equilibrium is closer to a Saha-Boltzman local thermodynamic equilibrium (LTE). This would imply that the temperature in the bombs is lower than the 80.000 K quoted above. Further aspects of this scenario are discussed in SM S3 and Fig. S7. In SM S5, we show that the absorption features indeed belong directly to the bombs.

The similarity of the profiles of C II, Si IV and Mg II in the bomb is quite striking. Normally, these lines show quite different shapes because of their different opacities and formation temperatures (Fig 2). In the bombs, however, a Si IV composite spectrum plotted over the C II and Mg II spectra shows a surprisingly similar overall shape (Fig. 3). Of course, C II and Mg II show strong self-absorption features in the line center. The self-absorption is slightly blueshifted because it originates from the overlying expanding chromosphere. But both the C II and the Mg II lines show a structure similar to Si IV farther away from the self-absorption center. Therefore, it is reasonable to assume that the plasma ejected in the explosion covers temperatures from 6000 K to 100,000 K. The high speeds exceeding 70 km/s are supersonic and close to or higher than the Alfvén speed. This supports the conclusion that we observe a multi-thermal reconnection outflow.

Energy is needed to accelerate, heat, and ionize the plasma when the bomb explodes. A rough estimate sets a lower limit for the required energy at $10^{22}$ J (=$10^{29}$ ergs), which must be dumped within, at most, a couple of minutes (SM S6). Thus for the brightest of our bombs the required energy exceeds estimates for traditional Ellerman bombs by an order of magnitude or more (*17*) and is a fraction of 0.1% to 1% of what is needed for a full-blown flare on the Sun that can reach up to $10^{32}$ ergs (*26,27*) – therefore the energy requirements for the bombs are substantial. The magnetic field strength near the surface in the regions where the bombs occur is almost 0.1 T, which means that drawing in the magnetic field from the surrounding volume and converting the magnetic energy in a reconnection process should deliver sufficient energy to power the bombs (*19*).

Because an observational link to Ellerman bombs could not be established yet, we cannot conclude if the bombs we report here are identical or different to the well-studied Ellerman bombs. Either they are a new class of energy release phenomena or they change our view of Ellerman bombs by showing that these reach much higher temperatures; given current energy estimations, both models and observations indicate temperature enhancements in Ellerman bombs of only a few 1000 K (*16,19*). Both conclusions mean that we have to revise our understanding of the dynamics and structure of the photosphere because dense material deep in the 6000 K cool near-surface regions can be heated to almost 100,000 K within minutes, which creates a temperature inversion in the solar atmosphere. Even modern three-dimensional time-dependent models fail to predict energetic events in which the dense plasma is heated to such high temperatures and accelerated to several times the speed of sound (*19*). These IRIS observations challenge our current view of the photospheric structure and dynamics and should stimulate further observational and theoretical research to investigate this interesting phenomenon.

HP kindly acknowledges the hospitality during a stay at LMSAL to get started with IRIS data. Warm thanks are due to Rob Rutten for his countless comments and for good discussions. IRIS is a NASA Small Explorer mission developed and operated by LMSAL with mission operations executed at NASA Ames Research center and major contributions to downlink communications funded by the Norwegian Space Center (NSC, Norway) through an ESA PRODEX contract. This work is supported by NASA contract NNG09FA40C (IRIS), the Lockheed Martin Independent Research Program, the European Research Council grant agreement No. 291058 and NASA grant NNX11AO98G. The SDO data used are provided courtesy of NASA/SDO and the AIA, and HMI science teams. All data used in this study are publicly available through the Virtual Solar Observatory (http://sdac.virtualsolar.org). The IRIS data are archived at http://iris.lmsal.com/data.html, where also manuals for data reduction are available. Sincere thanks are due to two anonymous referees for valuable comments.

**Supporting Material**

Supplementary Text S1 to S6

Figs. S1 to S9

Tables S1 to S2

Movie S1  ([http://www2.mps.mpg.de/data/outgoing/peter/papers/2014-iris-eb/fig1-movie.mov](http://www2.mps.mpg.de/data/outgoing/peter/papers/2014-iris-eb/fig1-movie.mov))

References (28-41) are called out only in the Supplementary Material.

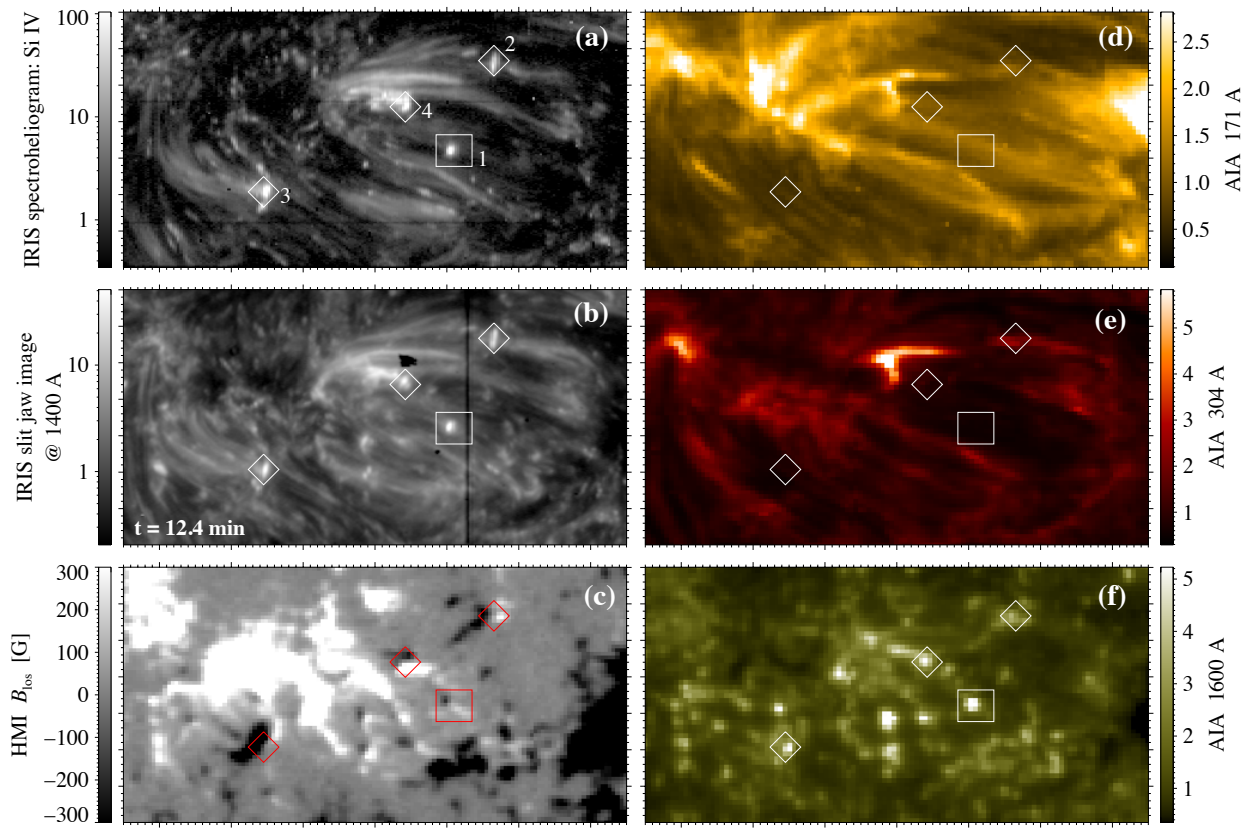

**Figure 1. Hot explosions, or bombs, in an emerging active region.**
Overview of the observations with IRIS and with other instruments providing context
from the photosphere into the corona. The field of view is 70 arcsec x 35 arcsec. (a) The
raster map in Si IV 1394 Å (80,000 K). The diamonds and the rectangle indicate the
bomb locations. (b) IRIS 1400 Å slit-jaw image dominated by Si IV. The vertical black
line is the slit. (c) The line-of-sight magnetic field at the surface obtained with the
Helioseismic and Magnetic Imager (HMI). The right panels show different channels of
the Atmospheric Imaging Assembly (AIA) with emission from (d) the corona at $10^6$ K.(e)
the transition region at $10^5$ K(f) the chromosphere below $10^4$ K. The intensity images are
normalized to the median intensity.
A movie showing the temporal evolution over 20 minutes is available online (Movie S1).

The movie is also available at
http://www2.mps.mpg.de/data/outgoing/peter/papers/2014-iris-eb/fig1-movie.mov.

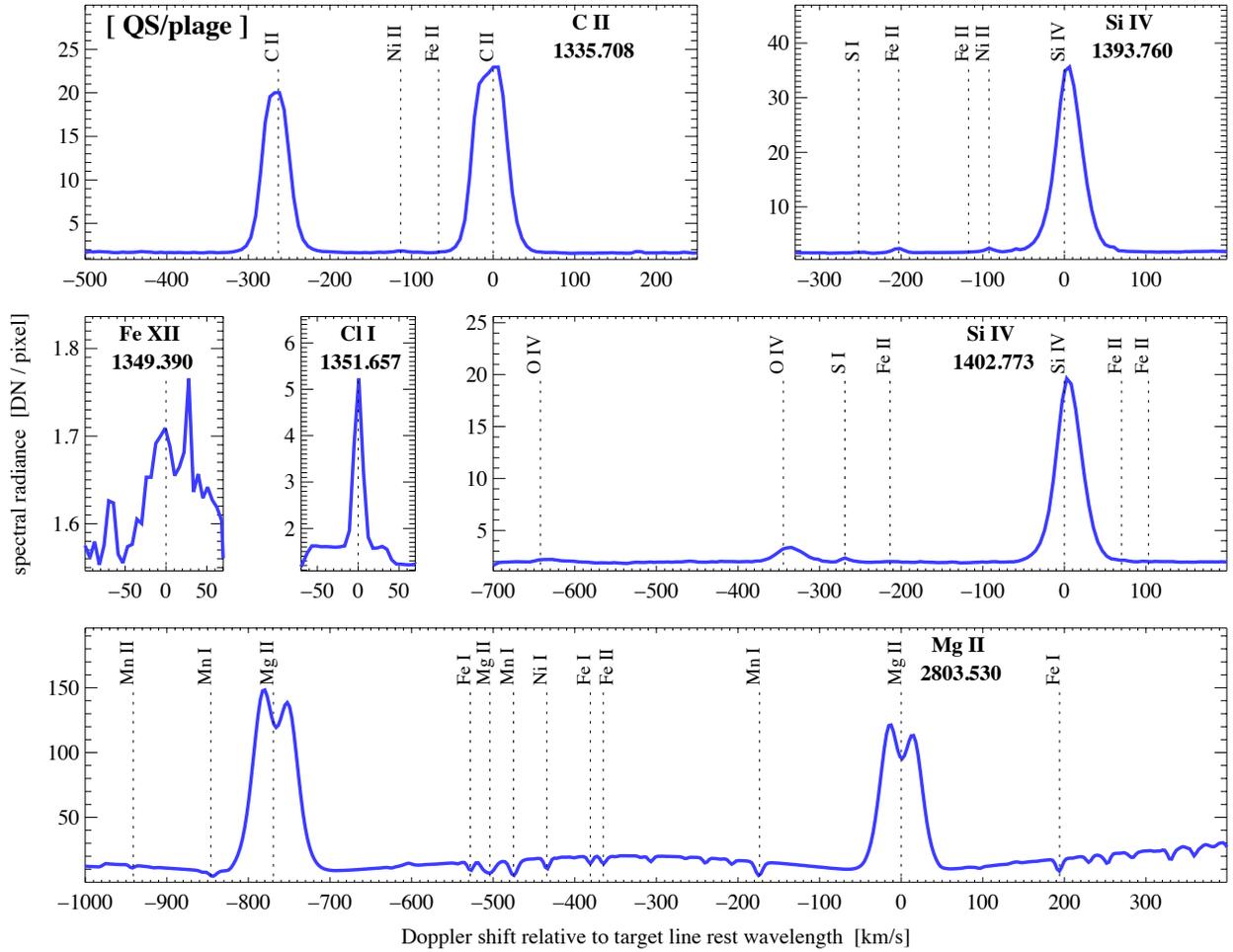

**Figure 2. Spectrum in the quiet Sun and plage area.**
This displays an average of the region surrounding the emerging active region showing spectral lines covering temperatures from 4000 K to 80,000 K (region marked in Fig. S1). The wavelength in each window is plotted in Doppler-shift units relative to the rest wavelength of the respective main target line (given in bold). The Doppler-shift scale in all panels is the same. The vertical dotted lines indicate the rest wavelengths for the (the main) spectral lines (see Table S1). This spectrum is averaged over a total area of about 37 arcsec x 110 arcsec and comprises about 75,000 individual spectra.

 

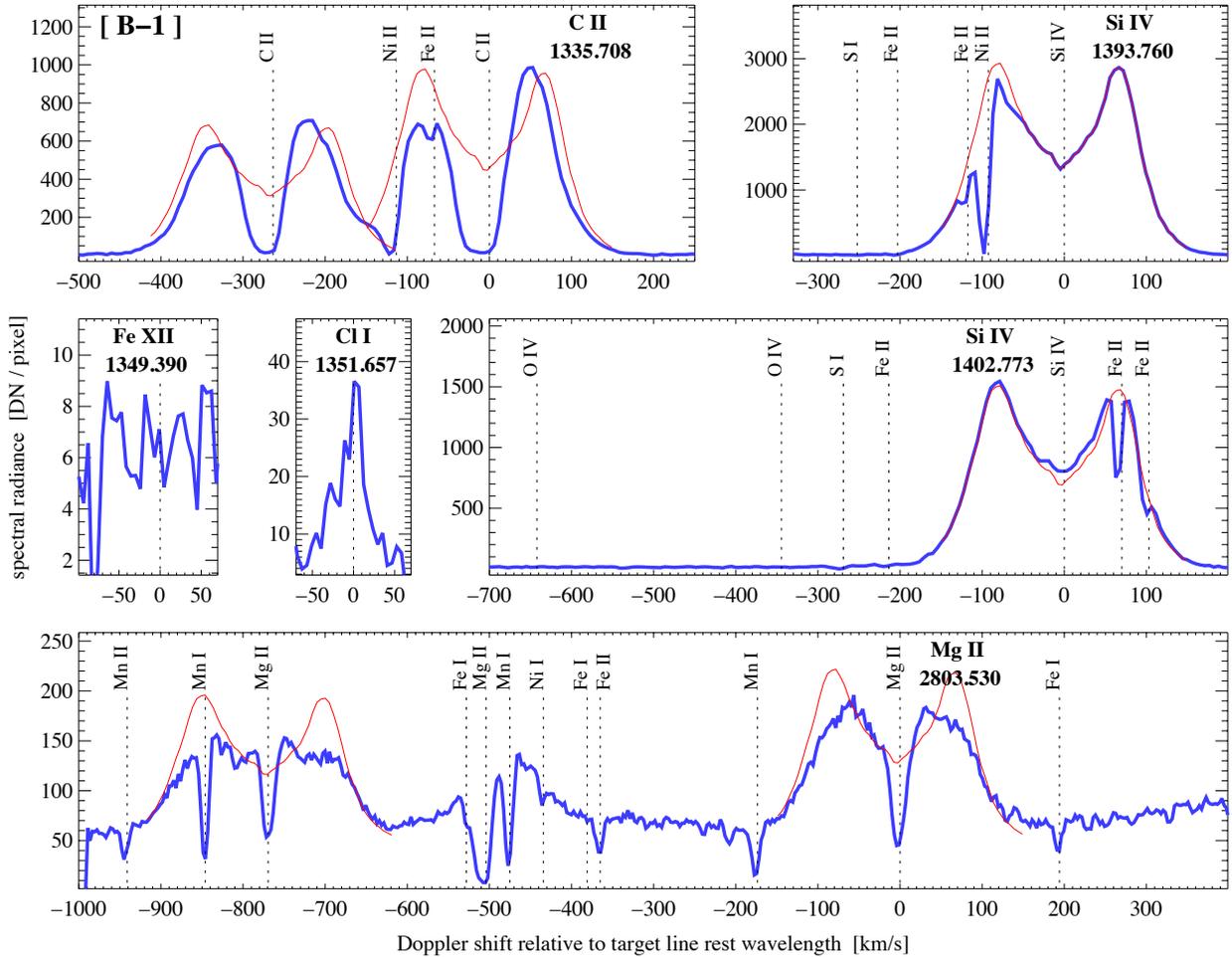

**Figure 3.  Spectrum of a hot explosion.**
High up- and downward velocities and cool overlying plasma can be seen in the spectrum of bomb 1 (in a single spatial pixel in the middle of the diamond marked 1 in Fig. 1). Wavelengths and labeling are identical to Fig. 2. The thin red line shows a Si IV composite spectrum plotted over the C II, Si IV, and Mg II lines (shifted to the respective rest wavelength and scaled in radiance to roughly match the C II and Mg II lines). The statistical errors in spectral radiance are $0.5 \times (DN/pixel)^{1/2}$ in the FUV channels (top and middle row) and $0.24 \times (DN/pixel)^{1/2}$ in the NUV channel (bottom row).

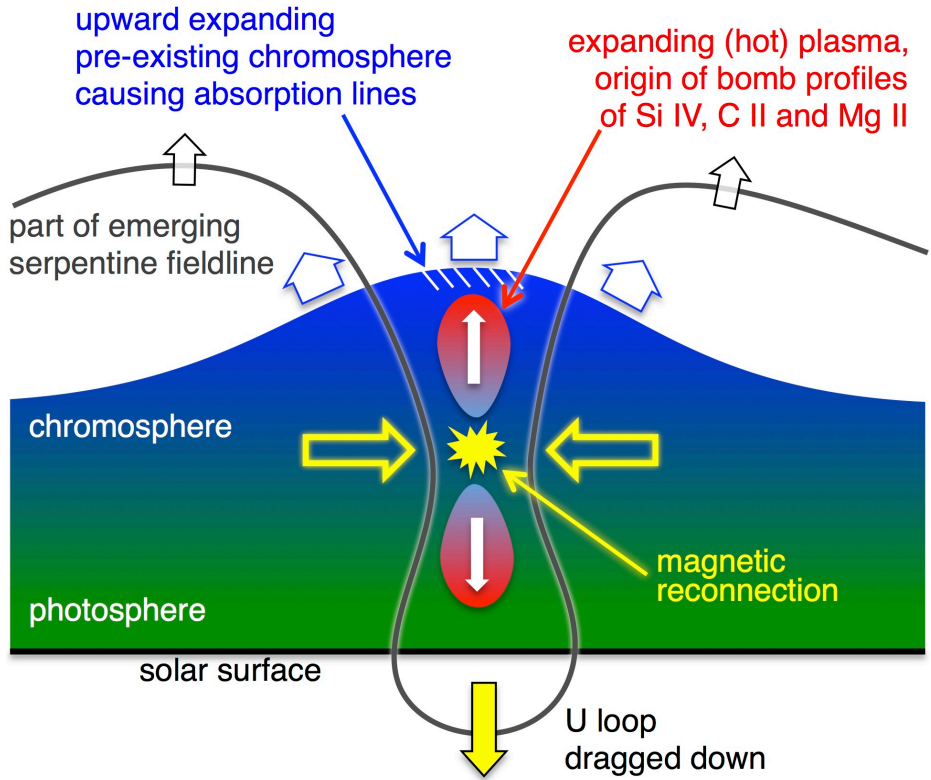

**Figure 4. Scenario for the hot explosions, or bombs.**
Cartoon of the bomb scenario. An undulating magnetic fieldline emerges and the resulting U-loop gets dragged down. Being squeezed together the magnetic field reconnects, and plasma is heated and accelerated deep in the atmosphere. The bidirectional outflow from the reconnection region causes the double-humped line profiles of Si IV, C II and Mg II, while the cool material above (white hashed area) causes the absorption lines.

Supplementary Materials for

# Hot Explosions in the Cool Atmosphere of the Sun

H. Peter,* H. Tian, W. Curdt, D. Schmit, D. Innes, B. De Pontieu, J. Lemen,
A. Title, P. Boerner, N. Hurlburt, T. D. Tarbell, J. P. Wuelser, Juan Martínez-Sykora,
L. Kleint, L. Golub, S. McKillop, K. K. Reeves, S. Saar, P. Testa, C. Kankelborg,
S. Jaeggli, M. Carlsson, V. Hansteen

*correspondence to:  peter@mps.mpg.de

**This PDF file includes:**

Supplementary Text S1 to S6
S1. Details of the Observations
S2. Temporal evolution of bombs, setting in the atmosphere, and maximum temperature
S3. Analysis of more bombs
S4. Density diagnostic in the bombs
S5. Possible filaments or fibrils as the cause of absorption lines?
S6. Estimating the energy requirements

Figs. S1 to S9
Tables S1 to S2
Caption for Movie S1

**Other Supplementary Materials for this manuscript includes the following:**

Movie S1:
http://www2.mps.mpg.de/data/outgoing/peter/papers/2014-iris-eb/fig1-movie.mov.

# Supplementary Text

## S1. Details of the observations

At the end of September 2013, IRIS performed several (identical) observing sequences to follow the emergence of an active region. In this study we selected one particular sequence that captured several brightenings with peculiar spectral profiles for a deeper investigation. Between 11:44 and 12:04 UT on 24 Sep 2013 IRIS acquired a raster scan with the spectrograph covering an area of 140 arcsec in the E-W direction and 175 arcsec in the N-S direction along the slit. The center of this raster map is at solar $(X, Y) = (-266$ arcsec, +83 arcsec), which corresponds to a heliocentric angle of 16 deg off disk center, i.e., close to disk center (the cosine of the angle is 0.96). A spectroheliogram in the Si IV line at 1394 Å is shown in Fig. S1. The spatial scale along the slit is 0.17 arcsec per pixel, corresponding to about 120 km on the Sun at disk center. The step size to produce the raster (E-W direction) is 0.35 arcsec. The exposure time for the individual spectra is 2 s. The IRIS obsID of this observing sequence was 4000254145. Calibrated level 2 data were used in our study, i.e., dark subtraction, flat field correction, and geometrical correction are taken into account (*9*).

The wavelength range covered includes several main target lines: C II at 1335 Å and 1336 Å, Si IV at 1394 Å and 1403 Å, the O IV lines at 1400 Å and 1401 Å, and Mg II at 2796 Å and 2804 Å. A more complete list of lines is given in Table S1. The spectral scale is 25 mÅ per pixel, corresponding to 5.5 km/s in Doppler units in the far UV channels (below 1500 Å) and 2.8 km/s in the near-UV channel (above 2700 Å), i.e., the data were binned by a factor of 2 in the spectral direction in the far-UV channel. Parts of the average spectrum are shown in Fig 2. The wavelength scale is converted into Doppler shifts to facilitate the interpretation in terms of flow speeds.

To perform the wavelength calibration, lines from neutral and single ionized species were used in the average spectrum of the quiet plage area surrounding the emerging magnetic region. These lines very likely show no noticeable shift, with the exception of C II and Mg II, which form under optically thick conditions in a more complicated way. Table S1 lists these lines in the far-UV region. In each wavelength window (separated by the horizontal lines) we also list the Doppler shift of the rest wavelength of a line with respect to the main target line in the respective window, labeled $\Delta v$. This facilitates navigating in the spectral plots. In the column titled plage we list the measured Doppler shifts of the lines. As expected, we find a net redshift of some 5 km/s to 10 km/s for the transition region lines of Si IV and O IV (*28*), which also confirms the wavelength calibration.

In addition to the spectral data, we used the slit-jaw images of IRIS that were recorded in the 1400 Å band. In the region of emerging flux these are dominated by the Si IV lines (with some contamination  from the continuum that forms near the temperature minimum). These images have a spatial scale of 0.17 arcsec per pixel, matching the slit spectra, and were recorded with a cadence of 12 s. They are easily aligned with the spectra using the fiducial marks on the slit.

To set the IRIS data in context, we used data recorded with the Solar Dynamics Observatory, SDO (*29*). In particular, we used filtergrams from the Atmospheric Imaging Assembly, AIA (*30*) in the 171 Å, 304 Å, and 1600 Å channels that mainly show emission from the million K corona in Fe IX, the transition region around 100,000 K in He II, and from the low chromosphere (with contamination from C IV), respectively. They have a spatial scale of about 0.6 arcsec per pixel and a cadence of 12 s, which matches the IRIS slit-jaw images (the 1600 Å channel has a lower cadence of only 24 s). We drew information on the magnetic field from the line-of-sight magnetograms recorded with the Helioseismic and Magnetic Imager, HMI (*31*). They have a spatial scale of about 0.5 arcsec per pixel and a cadence of 90 s. All these data were spatially aligned with the IRIS data and are further discussed in SOM S2.

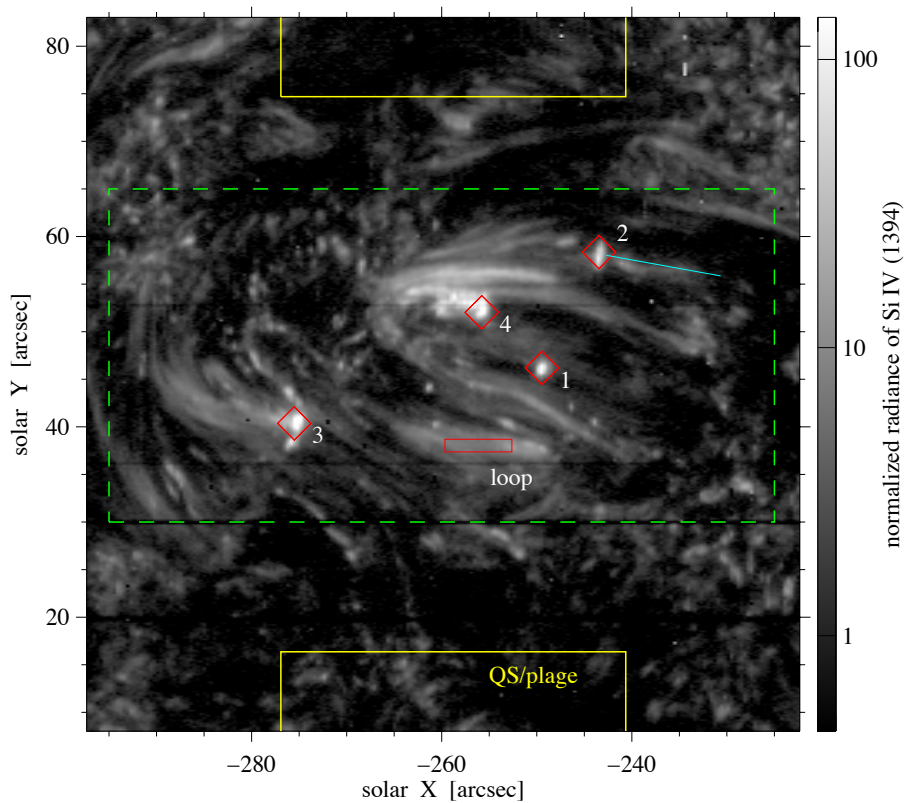

**Figure S1**. **Spectroheliogram of the emerging active region.**
This shows a 75x75 arcsec$^2$ part of the IRIS raster acquired in the spectral line of Si IV at 1394 Å. The red diamonds indicate the location of four roundish brightenings, bombs 1 to 4, captured during the raster. The red rectangle outlines the region of an emerging loop. The yellow boxes at the top and bottom indicate the part of the region used for wavelength calibration (extending 88 arcsec and 22 arcsec vertically for the top and bottom parts). It took about 10 minutes to acquire this part of the spectroheliogram. The green dashed rectangle outlines the region plotted in Fig. 1.

**Table S1. Lines of interest.**
Lines and their vacuum rest wavelengths in the far-UV band and their Doppler shifts in the quiet plage region and in the four bombs. The error in Doppler shift is about 1 km/s.

| line | wavelength at rest [Å] | $\Delta v$ [km/s] | Doppler shift [km/s] | | | | | Ref. |
|---|---|---|---|---|---|---|---|---|
| | | | plage | B–1 | B–2 | B–3 | B–4 | |
| C II | 1334.535 | −263.5 | −2.8 | [−10] | −20 | −35 | −55 | (S) |
| Ni II | 1335.203 | −113.4 | 0.0 | −6.5 | −7.2 | −6.8 | −4.0 | (S) |
| Fe II | 1335.409 | −67.1 | | −4.9 | | | | (N) |
| C II | 1335.708 | 0 | −3.4 | [−10] | −20 | −40 | −60 | (S) |
| Fe XII | 1349.39 | 0 | −4 | | | | | (S) |
| Cl I | 1351.657 | 0 | 0.0 | +1.5 | −1.0 | +0.2 | +0.4 | (S) |
| S I | 1392.589 | −252.1 | +0.3 | +0.1 | −3.6 | | | (S) |
| Fe II | 1392.816 | −203.2 | −0.5 | −3.9 | −7.6 | −9.9 | −3.6 | (S) |
| Fe II | 1393.214 | −117.5 | | −4.9 | | | | (K) |
| Ni II | 1393.33 | −92.5 | +0.6 | −5.7 | −5.8 | −6.6 | −2.0 | (S) |
| Si IV (1394) | 1393.76 | 0 | +5.9 | $\begin{cases}-81\\+66\end{cases}$ | −39 | −50 | −54 | (S) |
| O IV (1400) | 1399.775 | −642.1 | +10.7 | | | | | (S) |
| O IV (1401) | 1401.163 | −344.5 | +8.5 | | | | | (S) |
| S I | 1401.515 | −269.1 | +0.2 | −5.9 | | | | (S) |
| Fe II | 1401.774 | −213.7 | −1.9 | −6.1 | | | | (K) |
| Si IV (1403) | 1402.773 | 0 | +5.9 | $\begin{cases}-81\\+66\end{cases}$ | −30 | −42 | −47 | (S) |
| Fe II | 1403.101 | +70.1 | | −5.5 | | | | (K) |
| Fe II | 1403.255 | +103.0 | | −2.1 | | | | (K) |

*Notes to the table:*

- Lines that are not visible in the respective spectrum have no entry.

- All lines from neutral and singly ionized species (except for Cl I) are seen in *absorption* in the bomb spectra (B–1 to B–4). Those seen in absorption are marked in blue.

- The C II lines show a clear self-absorption in all bomb spectra, thus we list here some estimate for the Doppler shift based on the red and blue wing of the line.

- For the double-peaked Si IV profile in bomb 1 we list a separate value for the blue- and redshifted component.

- The values of $\Delta v$ list the Doppler shift of the rest wavelength of the respective line to the main target line in the wavelength windows (separated by horizontal lines) for better orientation in the plots of the spectra.

- References to rest wavelengths:

  (S): Rest wavelengths from the line list of Sandlin et al. (1986, ApJS 61, 801). These are based on solar observations (at the limb) and are all within 1 km/s (or 5 mÅ) of the theoretical values listed in Kurucz and NIST.

  (K): These lines have not been observed on the Sun in emission so far. The rest wavelengths adopted here are taken from the Kurucz line list (http://www.cfa.harvard.edu/amp/ampdata/kurucz23/sekur.html). They are within 1 km/s of the NIST data base. Ni II is not in the NIST list, but in the Kurucz list.

  (N): Taken from the NIST data base http://physics.nist.gov/PhysRefData/ASD/lines_form.html).

## S2. Temporal evolution of bombs, setting in the atmosphere, and maximum temperature

A snapshot of the region surrounding the four bombs is shown in Fig. 1. The temporal evolution of this is available as a movie in the supplementary material.

To investigate the evolution of the bombs we first investigated the change seen in the 1400 Å channel of IRIS that is dominated by the Si IV emission in active region areas. Because in the bombs the continuum as well as the line brightens up, the variation in the 1400 Å channel is a convolution of these two contributions. In plage areas the continuum level near the Si IV 1403 Å line is at about 2 counts per pixels (in the 2 s exposures), so it is very weak, barely above the noise level. The line *peak* of Si IV 1403 Å in those areas is about 20 counts (see Fig. 2). In the spectrum for bomb 1 the continuum intensity rises to about 20 counts and the Si IV 1403 Å peak to 2000 counts (and the line gets broader by a factor of almost 10, see Fig. S9b). So while during the bomb the continuum rises considerably by a factor of below 10, the line intensity increases even stronger, by a factor of 1000 (taking the broadening into account. Thus the time variation we see at the location of the bombs is indeed dominated by the variation in the Si IV lines.

In Fig. S2 we plot the variation of the slit-jaw-image intensity at the location of the bombs as a function of time. The bombs typically last for about 5 minutes and some show a signature of repetition. However, from the limited data sets we collected so far, we cannot conclude whether these time scales are related to the solar five-minute oscillations. The fact that the bombs last for about 5 minutes has an important implication. Because this is on the order of or longer than the ionization and recombination time of Si IV (*32*), ionization equilibrium can be reached for Si IV. Thus we can neglect non-equilibrium ionization effects when we explain the absence of the O IV lines in the bomb spectra (see SOM S4). We see some higher-frequency variability in the bomb lightcurves, which is similar to the lightcurves found in Ellerman bombs (*17*), which points to a dynamic nature of the reconnection process.

The HMI magnetograms, in particular their temporal evolution in the movie attached to Fig. 1, clearly show that the bombs are found above regions at the surface where magnetic flux of opposite polarity merges and cancels. To quantify this we display in Fig. S3 the temporal evolution of the magnetic flux in a rectangle around bomb 1, separately for the positive and negative flux. The short-term variations (over few minutes) are most probably due to small-scale non-resolved emerging and submerging flux or flux moving through the boundary of the rectangle. Considering longer time scales of 10 minutes or more, a clear trend can be seen. Both the positive and the negative flux are dropping, which shows that flux cancelation is ongoing in the photosphere below the bomb. This is an indirect sign of a reconnection process that might operate and deliver the (magnetic) energy needed to power the heating, ionization, and acceleration of the plasma. The pattern of merging opposite polarities is very clear for bombs 1 and 2 in the movie S1 attached to Fig 1, and less clear for bombs 3 and 4, because they are set in more complex magnetic environments.

The co-observations by AIA on SDO show that these bombs are also found in the chromosphere (1700 Å channel), but are not reflected in the hot coronal plasma (171 Å channel). Interestingly, they have no counterpart in the 304 Å channel either, which is dominated by emission from He II that forms at temperatures similar to Si IV. This absence of emission from the 171 and 304 channel (or of all the extreme UV channels of AIA, for that matter) is another property these bombs share with Ellerman bombs (*13*), even though we also found instance of bombs with some counterpart in the coronal channels. The absence or weakening of the extreme UV channels can be easily understood in terms of opacity. In the expanding overlying cool material above the bomb there will be a significant fraction of neutral hydrogen and helium and singly ionized helium. Therefore any light emitted at the wavelengths covered by the extreme-UV channels of AIA will be absorbed by the Lyman continua of H I (edge at 912 Å), He I (504 Å) and He II (228 Å).

The visibility of the bombs in Si IV implies that in those temperatures have to be reached that are close to the line-formation-temperature of Si IV. This is of the order of 80.000 K, but differs depending on how the line-formation-temperature is defined. Here we use the value following from an active region 3D MHD model including spectral synthesis (*32*). If one assumes that the line is formed near the peak of the ionization fraction of Si IV with or without considering the excitation process, or in a simple constant-pressure atmosphere, the line formation temperature would be lower by some 5% to 10% (*32*).

To determine whether material hotter than 100,000 K is produced in the bombs, one could check for an emission line formed at higher temperature with a wavelength longward of the Ly continua. In the IRIS range these would be Fe XII at 1349 Å, which forms at about 1.5 MK, and Fe XXI, which forms at 10 MK. The latter is seen only during flares. Unfortunately, the oscillator strength of Fe XII is too weak to be seen in the individual bombs. It is seen, however, when averaging over a large enough plage area (outside the bombs, cf. Fig. 2). Based on our current data, we therefore cannot say if the bombs become any hotter than 100,000 K because this would not be clearly visible in AIA coronal bands owing to the absorption by the cool material, and unfortunately, we do not have a strong enough line formed at a high enough temperature. This implies that our estimate of the maximum temperature reached in bombs is a lower limit.

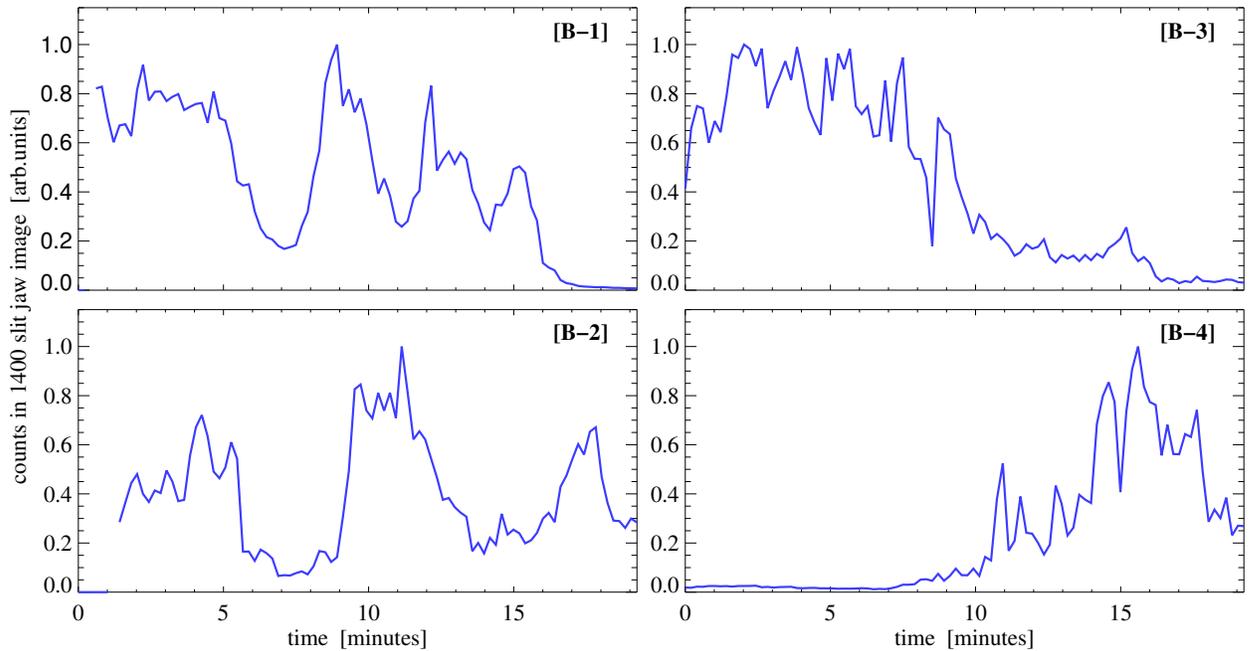

**Figure S2.  Temporal variation of the hot explosions.**
This shows the variation of the brightness of the four bombs as seen in the 1400 Å slit-jaw images of IRIS. We show the variation in the center pixel (0.17 x 0.17 arcsec$^2$) of the respective bomb (diamonds in Fig. 1). The statistical error in the slit-jaw count rates is about 1% to 2%.

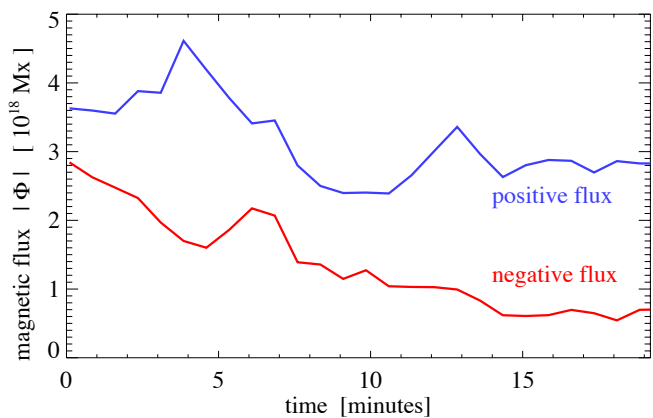

**Figure S3.  Temporal variation of magnetic flux below bomb 1.**
The magnetic flux is calculated separately for the positive and negative magnetic flux (white and black in Fig. 1) in the 7x6 arcsec$^2$ rectangle outlined in Fig. 1 around bomb 1.

### S3. Analysis of more bombs

In the main text we concentrated on bomb 1 and showed its spectrum in Fig 3. Now we discuss the other three bombs that are highlighted in Figs. 1 and S1 and compare them with the first case. Their spectra are plotted in Figs. S4 to S6. These other bombs also have a strongly enhanced intensity in Si IV and show no O IV lines. The main difference is that only bomb 1 shows a clear double peak in the Si IV profile that is indicative of a (vertical) bi-directional outflow, while the other three bombs show a single-Gaussian type strongly blueshifted profile.

Given the limited numbers of bombs studied so far, we speculate that this main difference basically is a line-of-sight effect. If the axis of the (reconnection) outflow is poorly aligned with the vertical, the down-flowing part of the reconnection outflow will be well hidden below the chromospheric material above it, as illustrated in Fig. S7). Thus any emission originating in the down-flowing part would be absorbed by the cool chromosphere above it through continuum absorption. In this interpretation the blueshifted emission profiles of bombs 2 to 4 would correspond to the blue hump in the profiles of bomb 1. Bomb 1 shows this nice double-humped profile only because the flow is aligned along the vertical line-of-sight. Only if the reconnection outflow is aligned with the line-of-sight one can look down to the redshifted downflow part through the heated blueshifted upflowing material (cf. Fig. 4). New studies including more bomb profiles will be needed to draw further and more solid conclusions.

Bomb 1 discussed in the main text is the richest in absorption lines from the cool material located above the hot explosion in the photosphere. Still, all the other bombs show absorption lines (see **Table S1**). The common pattern of all these absorption lines is that they consistently show blueshifts. This would be consistent with a slow upward motion of the photosphere and chromosphere during the flux emergence process (*24*). The only line from a neutral species that is *not* seen in absorption is Cl I at 1352 Å (cf. Figs. 3, S4, S5, S6). This line is optically pumped by the nearby C II lines (*33*). In particular, Cl I in the cool chromospheric layer above the bomb (cf. Fig. 4) will be pumped by the strong C II line originating from the hot bomb, and consequently will be seen in emission. Because Cl I originates from the cool chromosphere above the hot bomb, it will also be narrow, and not strongly broadened like C II or Si IV.

To compare the *overall shape* of the spectra of C II, Si IV and Mg II we calculated a composite Si IV profile. Because there are absorption lines in the blue half of the Si IV line at 1394 Å and in the red half of the line at 1403 Å, we simply combined the two halves of the spectra that are not (obviously) affected by absorption lines by scaling up the weaker of the Si IV lines by a factor of 1.95. We then plotted this composite Si IV profile in red on top of the C II, Si IV and Mg II lines after shifting it to the respective rest wavelength and scaling it in intensity so that it matches these lines. The ratio of the composite spectra of the Si IV lines at 1394 Å and 1403 Å is 1.95 in all cases (except for bomb 2, where it is 1.85). This is very close to 2, the theoretical value for optically thin conditions. We conclude that in all bombs Si IV forms under optically thin conditions.

Comparing the composite Si IV profile with C II and Mg II, we find that the latter ones show a profile very similar to Si IV, with the exception of bomb 2, when considering only the overall shape (not considering the self-absorption features). For bombs 3 and 4, where the Si IV profile is relatively close to a Gaussian but strongly blueshifted, we find that the self-absorption features of Mg II and C II are still close to the rest wavelength. This is consistent with the interpretation given for bomb 1, in which the strong enhancements of C II and Mg II form in the same reconnection outflow as Si IV, and then the self-absorption features originates from the expanding chromosphere overlaying the bomb. Bomb 2 does not show the close resemblance of the Si IV, C II and Mg II spectra, which might hint at a slightly different thermal structure for this event.

It has to be stressed that a detailed study of the formation of the Si IV, C II and Mg II lines in the framework of a realistic numerical 3D magneto-hydrodynamic model is needed to confirm these conclusions that are based on simple arguments. Unfortunately, the current numerical models either did not produce this type of bombs (or Ellerman bombs for that matter), or did not include an analysis of the expected spectral profiles for a detailed comparison with observations, as detailed in a recent review (*11*). Thus the IRIS observations we present here motivate new improved models of the dynamics of the solar photosphere.

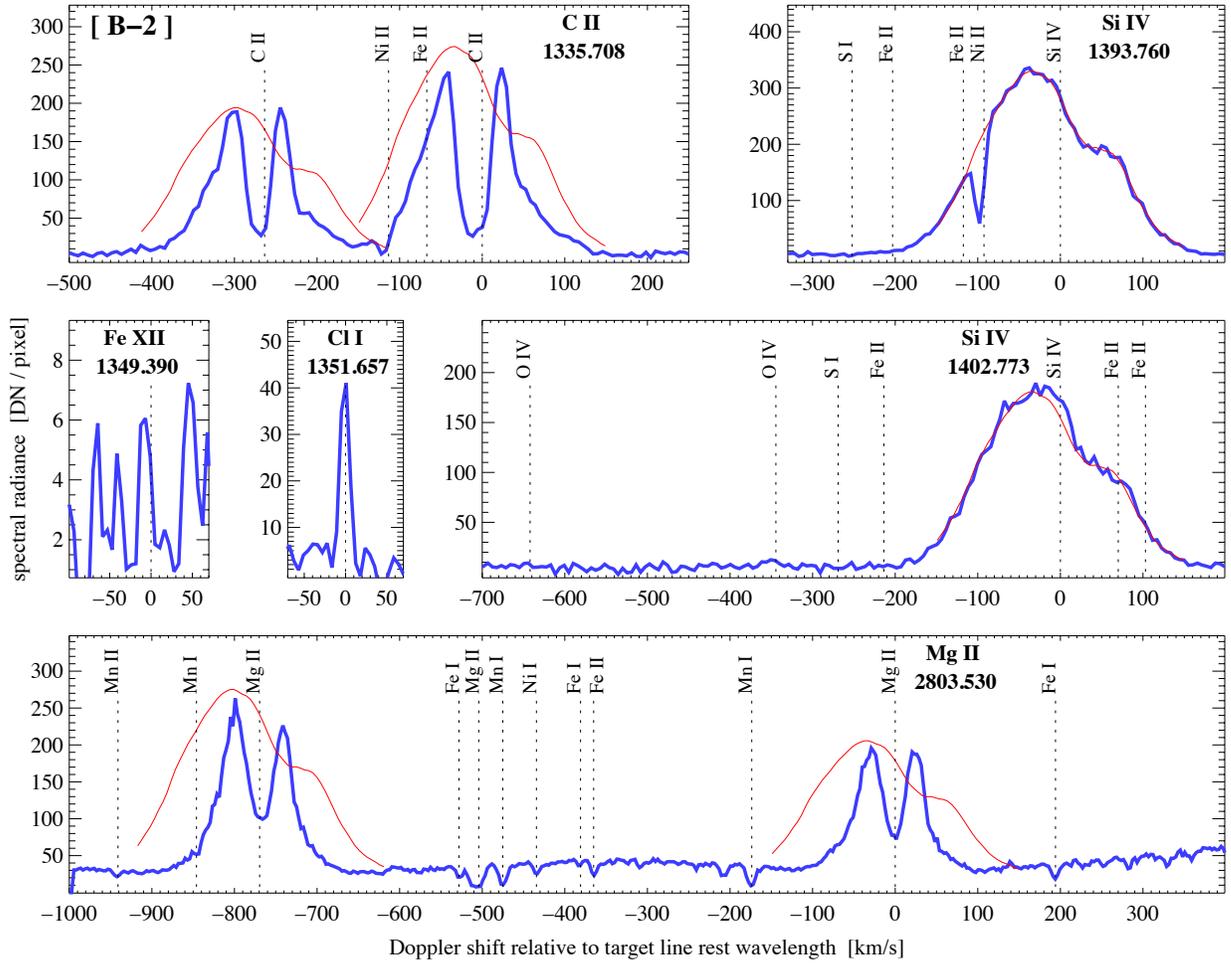

**Figure S4. Spectrum of bomb 2.**
Same as Fig. 3, but for a single spatial pixel at the location in the middle of the diamond marked 2 in Figs. 1 and S1. As in Fig. 3, the thin red line shows a Si IV composite spectrum overplotted on the main lines. The statistical errors in spectral radiance are $0.5 \times (DN/pixel)^{1/2}$ in the FUV channels (top and middle row) and $0.24 \times (DN/pixel)^{1/2}$ in the NUV channel (bottom row).

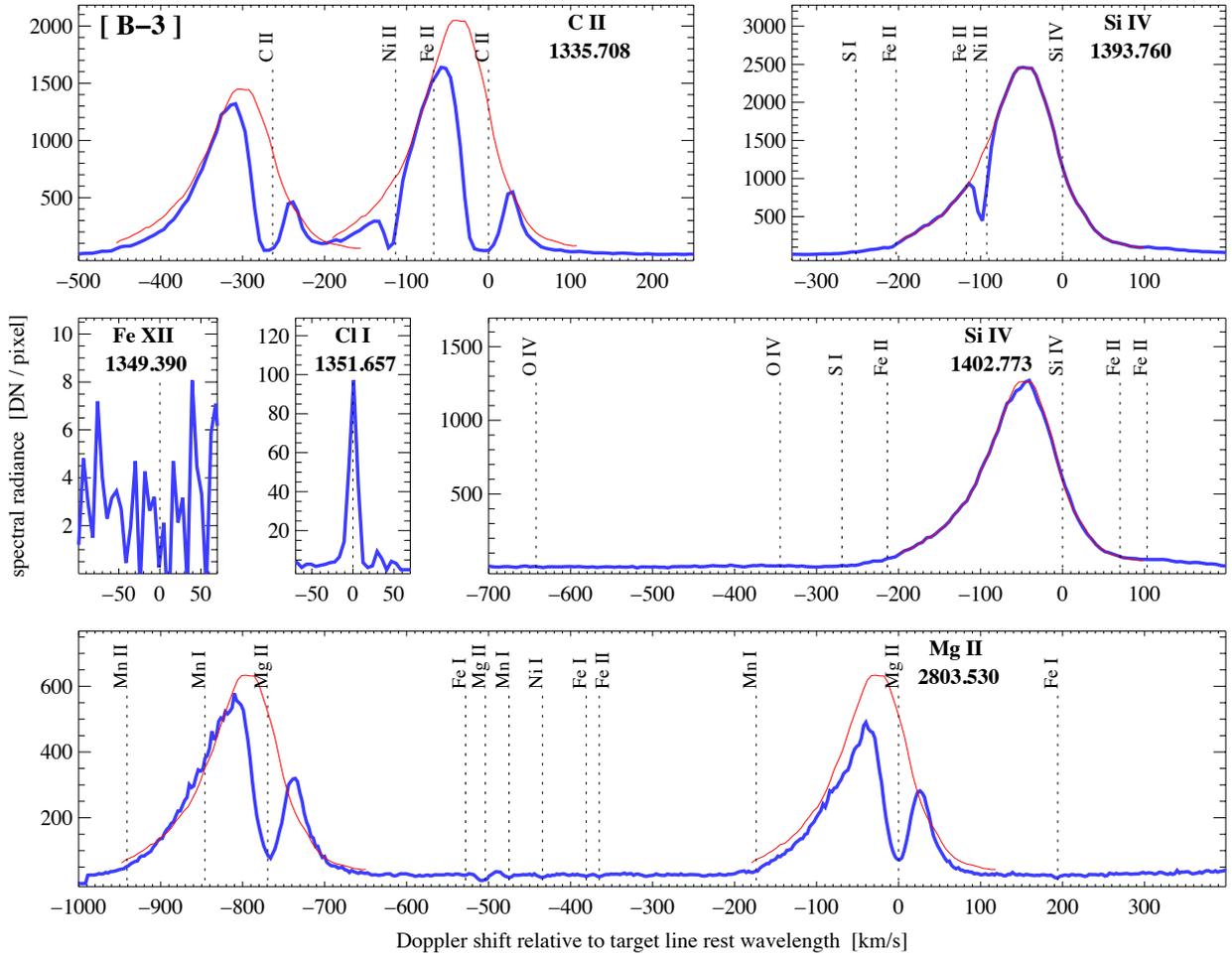

**Figure S5. Spectrum of bomb 3.**
Same as Fig. 3, but for a single spatial pixel at the location in the middle of the diamond marked 3 in Figs. 1 and S1. As in Fig. 3, the thin red line shows a Si IV composite spectrum overplotted on the main lines.

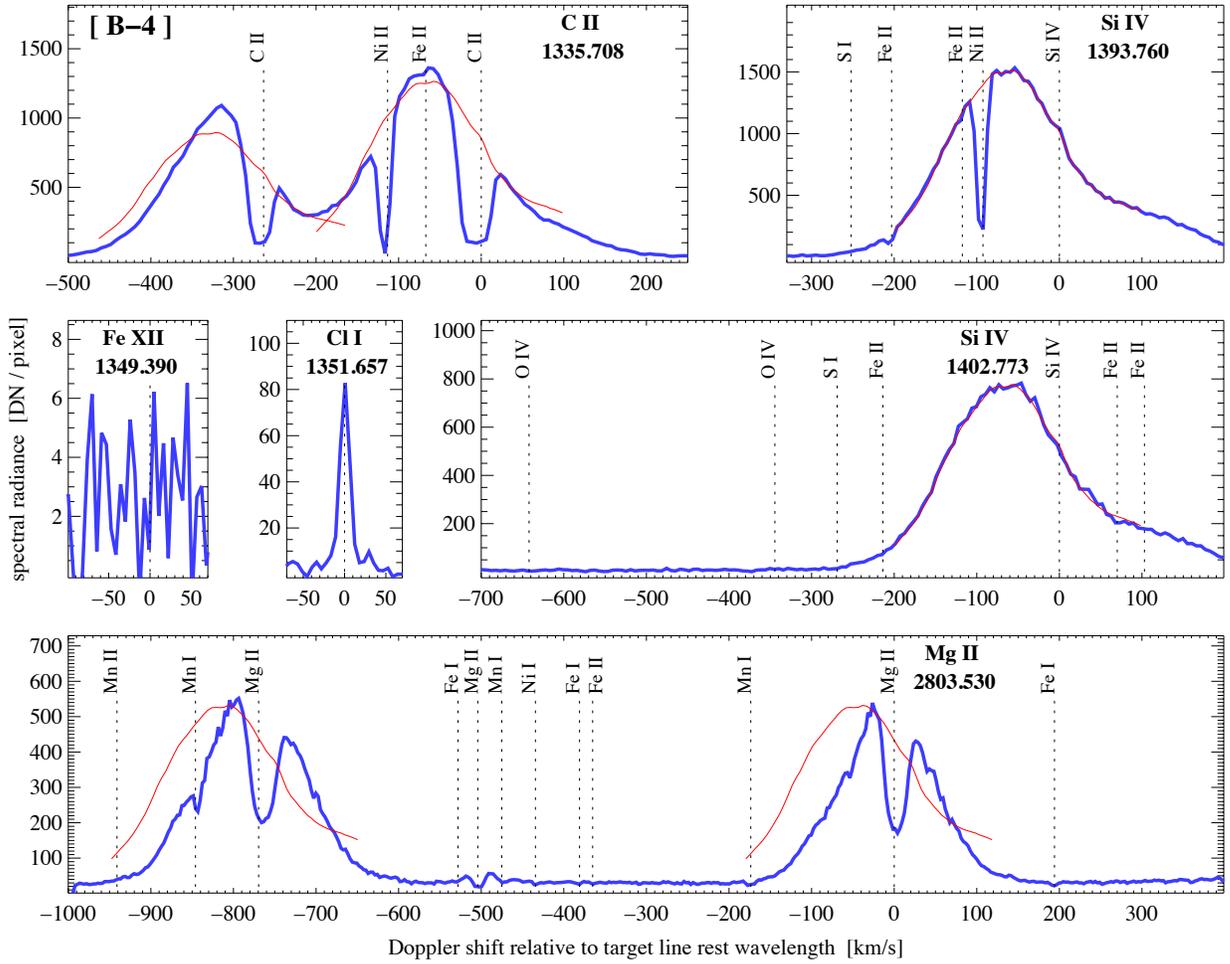

**Figure S6. Spectrum of bomb 4.**
Same as Fig. 3, but for a single spatial pixel at the location in the middle of the diamond marked 4 in Figs. 1 and S1. As in Fig. 3, the thin red line shows a Si IV composite spectrum overplotted on the main lines.

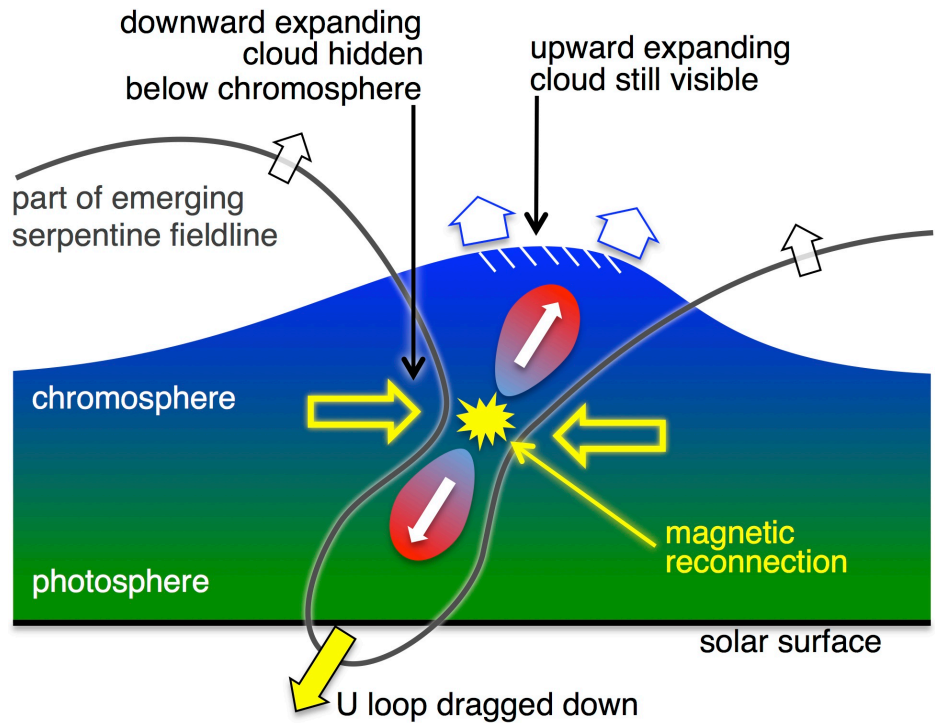

**Figure S7.  Scenario for an inclined bomb.**
This is similar to Fig. 4, but now for the case if the reconnection outflow is not aligned with the vertical. Then the downward flowing part of the reconnection outflow is hidden below the thick chromosphere and only the upward flowing part is visible. Hence the line profiles will show only one hump in the blue instead of a symmetric double-humped profile as for the vertical bomb shown in Fig. 4.

## S4. Density diagnostics in the bombs

The ratio of the forbidden O IV lines in the IRIS wavelength range can be used to investigate the density in the source region of these lines (*34*). However, the O IV lines are absent in the bombs, which precludes this type of density diagnostics and indicates that the density is high enough for collisional de-excitation from the metastable level to be more efficient than radiative decay. Still, a lower limit for the electron density can be derived in the following way: with increasing electron density the de-excitation of the O IV lines due to electron collisions will become increasingly efficient. This means that above a certain threshold the O IV lines will no longer be visible. The lower limit for the ratio of the Si IV line at 1394 Å to, e.g., the O IV line at 1401 Å will thus provide a lower limit of the electron density.

For this a theoretical ratio of these two lines as a function of density is needed. For simplicity we assume that the source regions of the O IV and Si IV lines are in pressure equilibrium and that the lines form at their respective peak of the contribution function. Then the ratio depends on the electron density as shown in Fig. S8. Because we assume pressure equilibrium, we find an electron density at the source regions of both O IV and of Si IV, which are labeled at the top and bottom axis in the figure. These calculations were made using CHIANTI version 7.1.3 (*35*) with the photospheric abundances and the Mazzotta ionization equilibrium as included in CHIANTI. The threshold at low densities (below ~$10^{10}$ cm$^{-3}$) depends on the line formation temperature assumed for Si IV and O IV. To a lesser degree this also applies to the relation of density and line ratio at higher densities. For consistency we choose here the temperature of the peak of the respective ion fractions based on the data in CHIANTI ($\log_{10} T$ [K] = 4.87 and 5.19 for Si IV and O IV).

To test this method, we used the spectrum from the region with the emerging loop shown in Fig. S9f (the location of that loop is indicated in Fig. S1 by a rectangle). Here the ratio of the O IV can be used in the classical way to retrieve the electron density (at the source region of O IV). The ratio of the 1401 Å line to the 1400 Å line of O IV is 0.26. Using the standard tools available with CHIANTI, the electron density in the source region of O IV is about $\log_{10} n_e$ [cm$^{-3}$] = 10.7. For the line ratio of Si IV at 1394 Å to O IV at 1401 Å we find a ratio of about 10. Reading then the electron density at O IV off the graph in Fig. S8, we find an electron density at the source region of O IV of $\log_{10} n_e$ [cm$^{-3}$] = 11.3. This is within a factor of 4 of the traditional method, which we consider sufficient for the purpose at hand to get an order-of-magnitude estimate.

In Fig. S9 we plot in the four middle panels (b–e) the part of the bomb spectra that covers the O IV line at 1401 Å, now on a logarithmic scale, to highlight the details at low count rates. As before, we overplot the composite Si IV profile, but now also at the location where the O IV line would be. We scale the intensity such that the scaled composite profile would be the strongest possible O IV profile still compatible with not being seen above the continuum (cf. SOM S2). Of course, this only provides a rough lower limit for the Si IV to O IV ratio, but that is sufficient for this estimation. We list these ratio limits in Table S2. Reading the corresponding electron densities at the source region of Si IV off the graph in Fig. S8, we list these in Table S2, as well. Clearly, the electron densities in the source region of Si IV are generally on the order of $10^{13}$ cm$^{-3}$, or even higher. In particular, for bomb 1 the lower limit of the *electron* density at the source region of Si IV (at a temperature of about 80,000 K) would be about $5 \times 10^{13}$ cm$^{-3}$.

In the hot pockets where we find Si IV we can assume the plasma to be fully ionized. (In equilibrium hydrogen is 99% ionized already at 30.000 K; *36*). Ionization equilibrium can be reached because the bombs last for about five minutes, while the ionization time of hydrogen is only of the order of one minute. Therefore the *hydrogen* (proton) density in the Si IV source region will be similar to that of the electrons. Now the bomb is also seen in the Mg II line, and Mg II will be confined to cool regions below 15,000 K (*37*), otherwise the ion would quickly be further ionized. We assume that the source regions of both Mg II and Si IV are magnetically connected to the energy deposition site. Thus the source regions of Mg II and Si IV are connected and we can assume that these two regions are in pressure equilibrium (along the magnetic field). This is also supported by the similarity of the overall shapes of the profiles of these two lines. Consequently the density in the Mg II source region should be higher by the ratio of the line formation temperatures, i.e. about 80,000/15,000. So we derive a lower limit for the *hydrogen* density in the Mg II source region on the order of $3 \times 10^{14}$ cm$^{-3}$. According to semiempirical models, this density corresponds to the temperature minimum at some 500 km above the surface. Because we estimate only a lower limit for the density, the bombs will be set at higher densities at deeper layers, i.e., in the photosphere.

An alternative explanation of the missing O IV lines would be to assume a non-Maxwellian distribution function of the electrons. With kappa-distributions one can construct spectra with weak O IV emission at lower densities as well (*38*). Electron beams might give rise to such non-Maxwellian distributions, for instance. IRIS observations indeed provide evidence for non-thermal electrons in active regions (*39*). However, together with the other evidence of the bombs taking place in the low atmosphere, this option seems unlikely.

If the temperature would be strictly limited below 80,000 K, one would also expect a reduction of the O IV emission that in equilibrium forms at 150,000 K. However, this would explain only a reduction by a factor of about 10, and not O IV being completely absent. Thus this effects could only be part of the story.

Another consideration would be the change of abundances during the process. It is well known that the abundances change from the surface into the wind and preferentially bring elements with a low first-ionization potential (FIP) into the wind. However, this FIP effect (*40*) changes the abundances by no more than a factor of 3, therefore this can be neglected because of the dramatic weakening of the O IV lines.

Concluding from this, the most likely scenario for the absence of the O IV lines is due to high density in the bombs, reaching up to $3 \times 10^{14}$ cm$^{-3}$ in the source region of Mg II.

**Table S2.  Density diagnostics in the bombs.**
Lower limits for the Si IV to O IV line ratios in the four bombs and lower limits for the electron density deduced from this. We do not specify error estimates for the line rations (and the derived densities), because the ratios are rough lower limits only.

|  | Bomb 1 | Bomb 2 | Bomb 3 | Bomb 4 |
|---|---|---|---|---|
| ratio  Si IV 1394 Å / O IV 1401 Å | > 700 | > 100 | > 200 | > 400 |
| electron density at source region of Si IV $\log_{10} n_e$  [cm$^{-3}$] | > 13.7 | > 12.8 | > 13.1 | > 13.4 |

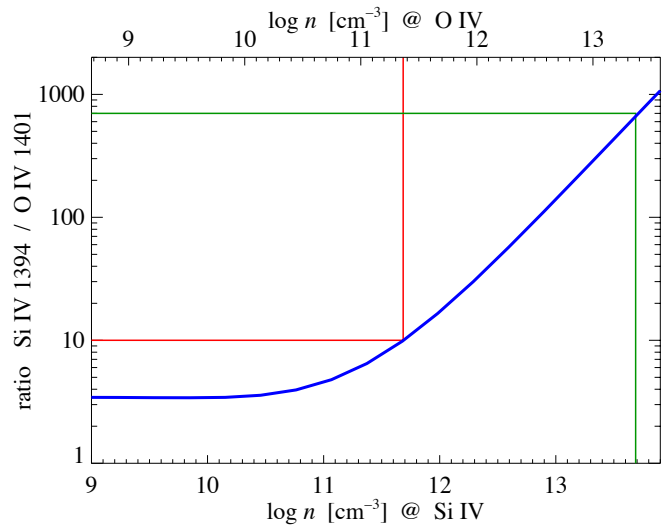

**Figure S8.  Density diagnostics from the ratio of Si IV to O IV.**
Predicted ratio assuming that both lines form at the same pressure. The straight lines indicate the ratios and densities associated with the lower limits of Si IV to O IV derived for the emerging loop (red) and bomb 1 (green) as listed in Table S2.

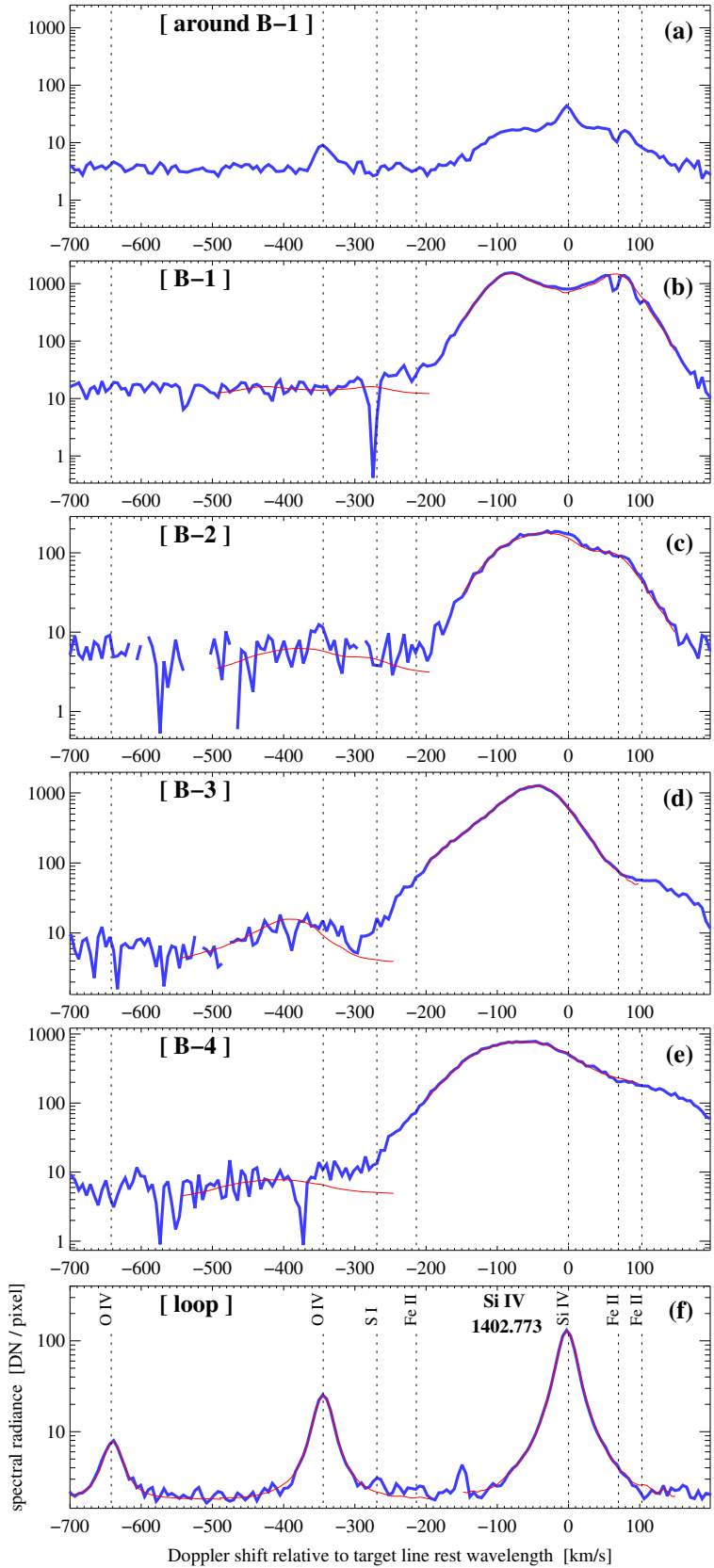

**Figure S9.**
**Comparison of bomb spectra.**

Spectra on logarithmic scale to highlight emission at low count rates. They show the same Si IV 1403 Å windows as in the other figures with the spectra on a linear scale.

The bottom panel shows the spectrum of an emerging loop marked in Fig. S1 by a rectangle.

The middle four spectra show the bomb spectra and are identical to those in Fig. 1 and Figs. S4 to S6, except that they are on logarithmic scale.

The top spectrum shows the average spectrum around bomb 1.

As before, the red lines show the composite Si IV spectrum, scaled to match the 1403 Å line and placed at the O IV 1401 Å position to estimate a lower limit for the Si IV-to-O IV ratio.

The statistical errors in spectral radiance are $0.5 \times (DN/pixel)^{1/2}$.

### S5.  Possible filaments or fibrils as the cause of absorption lines?

Discussing the bomb scenario in the main text and in SOM S3 we imply that the source region of the absorption features seen in the bomb spectra are directly linked to the bombs and the related flux emergence. One might also speculate that filaments and fibrils, i.e., cool structures that can be magnetically trapped in the higher (hot) atmosphere, could be responsible for these absorption lines. Still, this would not challenge our conclusion that the bombs are located in the photosphere, which is also based on the high density of the plasma in the bomb.

The bombs are very compact objects, with diameters smaller than 2 Mm. In contrast, (parts of) filaments or fibrils are lengthy objects with similar (or smaller) diameters, but much greater lengths. If the absorption were to be caused by a fibril, for example, this absorption feature would be detected not only directly above the bomb, but also to at least one side of the bomb.

A well-suited line for such a test is S I at 1402 Å, which has a rest wavelength of about 270 km/s blueward of the Si IV 1403 Å line. This is because it is in the continuum and thus its presence as an absorption line does not depend on the presence of the enhanced Si IV profile in the bomb. In the average spectrum it is seen in emission (cf. Fig. 2). In the spectrum of bomb 1 it is seen in absorption. This is very clear when plotting the spectrum on a logarithmic scale (Fig. S9b). But we do not clearly see this absorption feature in the vicinity of the bomb. To illustrate this, we plot the spectrum averaged around the bomb in a rectangular shell (this is the average along the one-pixel wide circumference of a rectangle with size 1.75 arcsec x 2 arcsec around the bomb, Fig. S9a). This is still not quite big enough to fully encircle the bomb, because the enhanced broad emission of Si is still visible. Nonetheless, the S I absorption line is no longer visible, even though the continuum level would be high enough to detect it (the bomb spectrum and the spectrum around it, i.e., the top panels a and b of Fig. 8, are plotted on the same intensity scale). While we concentrated here on the S I absorption feature (which is visible only in bomb 1), the arguments hold in the same way for the other absorption features from Ni II and Fe II seen in the Si IV lines. While we do not see the absorption features clearly in the places outside the bombs (the signal-to noise becomes an issue here), we cannot rule out their presence. Still, if present around the bomb, the absorption feature is much weaker than directly above the bomb.

From this consideration and because all absorption lines are blueshifted we safely conclude that the absorption lines seen in the C II and Si IV are primarily due to chromospheric material directly above the explosion that is lifted in the flux-emergence process.

## S6. Estimating the energy requirements

We first estimate the energy required to heat and accelerate the hot plasma in the source region of Si IV and the consider the Mg II source region. This is merely an order-of-magnitude estimation.

In general, the thermal energy density is given by $e_{th} = 3/2\, n\, k_B\, T$, where $n$ is the particle density, $k_B$ is the Boltzmann constant, and $T$ is the temperature. If the flow speed is $v$, the kinetic energy is given by $e_{kin} = \frac{1}{2}\, n\, \mu\, v^2$, where the mean molecular weight $\mu$ can be assumed to be about 0.6 of the proton mass (if the plasma is fully ionized). Finally, the energy needed for ionization needs to be accounted for, $e_{ion} = n\, E_H$, where $E_H$=13.6 eV is the ionization potential of hydrogen. Assuming that the plasma became ionized from being neutral, here $n$ would be the same as the electron density. Finally, the total energy requirement is given by $E_{Si\,IV} = V\,(\,e_{th} + e_{kin} + e_{ion}\,)$, where $V$ is the volume of the source region.

For the temperature we assume that the plasma has been heated to $T = 10^5$ K to emit Si IV, and for the density we assume n = $5\times10^{13}$ cm$^{-3}$ as derived in S4 for the source region of Si IV. In bomb 1 we see the bi-directional flow with $v = 75$ km/s in both directions, so we use this value here. For the volume $V$ we assume that the bomb is a cube with the side-length equal to the horizontal extent of the bombs of about 2 Mm, which is a rough estimate only.

With these values we find that roughly the same amounts of energy are needed to heat ($e_{th}$), accelerate ($e_{kin}$), and ionize ($e_{ion}$) the plasma in the source region of Si IV. Together, the total energy requirement for the source region of Si IV then is about $E_{Si\,IV} = 5\times10^{21}$ J (or $5\times10^{28}$ ergs).

However, considering also the Mg II source region this is an underestimation. As mentioned in the main text, in the bombs the Mg II lines show profiles that are (outside the self-absorption features) very similar to Si IV. Therefore it might be that the outer parts (in the wing) of Mg II are a reflection of the reconnection outflow, indicating a strong flow with a magnitude similar to what is also seen in Si IV in the cooler plasma where Mg II originates. Assuming the density estimated for the Mg II region in SOM S4, i.e., $3\times10^{14}$ cm$^{-3}$, we then find that almost $10^{22}$ J (or $10^{29}$ ergs) would be needed to power the flow in the photosphere (here there is no significant need to ionize or heat the plasma). This is a substantial fraction of the energy needed to power a full-blown solar flare, for which typically up to $10^{25}$ J (or $10^{32}$ ergs) are required – and we did not yet consider the energy required to power the radiative losses. Of course, these values should be taken cum grano salis, because detailed numerical modeling is required to do more solid estimates.

The energy to heat the bomb to 100,000 K and to power the flow in the photosphere can be drawn from converting magnetic energy. The magnetic energy density is given by $e_{mag}=B^2/(2\mu_0)$, where B is the magnetic field strength and $\mu_0$ the magnetic permeability. Assuming $B$=0.05 T (cf. Fig. 1) in the bomb, which is close to the value in the merging bipole at the surface, in the volume of the bomb the magnetic energy $E_{mag}=V\,e_{mag}$ is about $2\times10^{22}$ J. This is about twice as much as needed to heat and accelerate the plasma. Of course, one cannot convert all the magnetic energy. In flares the estimations range from 5% to 20% of the total magnetic energy (*41*), but they might be less efficient than the compact small-scale bombs. If one assumes that the magnetic field is drawn in from the sides while the reconnection process is underway, the volume from which magnetic energy is drawn can be larger, so that in the end there should be a sufficient supply of energy to power the bombs described here.

In principle, when converting magnetic energy (fully) into kinetic energy, the resulting velocity should be the Alfven speed. This is consistent with the bi-directional flow speed of some 75 km/s being comparable to the Alfven speed. Converting magnetic energy to thermal energy, the temperature increase should correspond to plasma-$\beta$, which is the ratio of plasma pressure to magnetic pressure. With the numbers above, plasma-$\beta$ would be about 1/40, being sufficient for a temperature rise from 6000 K to 100,000 K. Again, these energy estimates can be considered as a first guess only, because as of yet there are no realistic numerical models for these bombs, just as for their long-studied relatives, the Ellerman bombs.

**Movie S1**
**Temporal evolution of the emerging active region.**

A still of this movie is shown in Fig. 1. Panel (a) shows the raster map in the Si IV 1394 Å (80,000 K). Panel (b) shows the IRIS 1400 Å slit-jaw image dominated by Si IV. The vertical black line is the slit. The line-of-sight magnetic field at the surface obtained with HMI is displayed in panel (c). The right panels show different AIA channels with emission from the corona at $10^6$ K (d), the transition region at $10^5$ K (e), and the chromosphere below $10^4$ K (f). The field of view is 70 arcsec x 35 arcsec.

The movie is available at
http://www2.mps.mpg.de/data/outgoing/peter/papers/2014-iris-eb/fig1-movie.mov.